\documentclass[aps,pra,twocolumn,showpacs,superscriptaddress,groupedaddress]{revtex4-1}  % for review and submission
\usepackage{graphicx}  % needed for figures
\usepackage{dcolumn}   % needed for some tables
\usepackage{bm}        % for math
\usepackage{amssymb}   % for math
\usepackage{amsmath}   % for math
\usepackage{amssymb}   % for mathbb

\begin{document}

\title{EIT-Enhanced Coupled-Resonance Spectroscopy}
\author{Junyi Zhang}
\email{junyiz@princeton.edu}
\affiliation{Department of Physics, Princeton University, Princeton, 08544, New Jersey, USA }

\date{\today}

\begin{abstract}
Coupled-resonance spectroscopy has been recently reported and applied for spectroscopic measurements and laser stabilizations.  With coupled-resonance spectroscopy, one may indirectly measure some transitions between the excited states that are hard to be measured directly because of  the lack of populations in the excited states.
An improvement of the coupled-resonance spectroscopy  by combining the technology of electromagnetically induced transparency (EIT) is proposed.  The coupled-resonance spectroscopy signal can be significantly enhanced by EIT.  
Several experimental schemes have been discussed. 
The line shape of the EIT-enhanced coupled-resonance spectroscopy has been calculated.
The EIT-enhanced coupled-resonance spectroscopy can be used for simultaneously stabilizing two lasers to the same atomic source.  
\end{abstract}

\maketitle
\section{Introduction}\label{sec:Intro}
High precision spectroscopic measurements and laser stabilizations provide fundamental technologies for various modern physics researches and applications, such as atomic clocks~\cite{Hinkley2013,Bloom2014}, gravitational wave detection~\cite{Kolkowitz2016}, precision measurements of fundamental constants~\cite{Rosenband2008}, and searching for new physics beyond the standard model~\cite{Blatt2008}.

The thermal motion of the atoms causes Doppler broadening that is several orders of magnitudes larger than the natural line with at room temperature. So the Doppler-free saturation spectroscopy~\cite{Bennett1962,Lamb1964} has been widely used for spectroscopic measurements and for locking lasers to the atomic transitions.  Although it has some variants~\cite{Wieman1976} it usually involves a pair of counter-propagating pump and probe beams.  The pump beam saturates the transition, which leads to an Doppler-free dip of absorption near the center of the Doppler-broadened absorption.

However, the direct spectroscopic measurements of the transitions between the excited states are more difficult because of the lack of the thermal population.  Recently, several schemes of coupled-resonance spectroscopy have been proposed \cite{Burd2014, Barker2015, Zhang2016}, where insufficiently populated transition $C$ is coupled indirectly (spontaneous emission)  or directly to another transition $P$ that can be more easily measured, so that the the information of transition $C$ can be probed from $P$.

The coupled-resonance spectroscopy of $\text{Yb}^+$ was reported in Ref.~\citenum{Burd2014},  where the transition $\ ^2D_{3/2} \rightarrow \ ^3D[3/2]_{1/2}$ ($935 \text{nm}$)  is measured by coupling it to the transition $\ ^2S_{1/2} \rightarrow \ ^2P_{1/2}$ ($369 \text{nm}$) .  The upper states $\ ^3D[3/2]_{1/2}$ and $\ ^2P_{1/2}$ can spontaneously decay into both the lower states $\ ^2S_{1/2} $ and $ \ ^2D_{3/2}$, which forms a four-level closed loop.  The experimental results agree with the rate equations. 
In Ref.~\citenum{Barker2015} and \citenum{Zhang2016},  coupled-resonance spectroscopy signal of the transition $\ ^3P_{0} \rightarrow \ ^3S_{1}$ ($679 \text{nm}$)   of $\text{Sr}$  was observed by coupling to the transition $\ ^3P_{2} \rightarrow \ ^3S_{1}$ ($707 \text{nm}$). Since the both transitions share the same upper state, it forms a $\Lambda$-scheme in contrast to Ref.~\citenum{Burd2014}. The decay rate of the transition $\ ^3P_{0} \rightarrow \ ^3S_{1}$ is $1.5\text{MHz}$, about only $1/5$ of the $\ ^3P_{2} \rightarrow \ ^3S_{1}$ and both lower states are excited states, so the signal is very weak.

Here, an improvement of coupled-resonance spectroscopy for the directly coupled schemes is proposed. If the coupling transition is coupled coherently to the probe transition, the coupled-resonance spectroscopy signal can be enhanced by the quantum interference. Similar to the electromagnetically induced transparency (EIT)~\cite{Harris1990,Fleischhauer2005} , the linear susceptibility may vanish at the two-photon resonance, therefore we refer to it as EIT-enhanced coupled-resonance spectroscopy. The EIT-enhanced coupled-resonance spectroscopy is not limited to the $\Lambda$-scheme, it can also be  applied to the ladder- and $V$-schemes.       

The rest of this article is organized as follows.  In Section~\ref{sec:Experiment}, the level schemes are discussed and qualitatively compared to conventional coupled-resonance spectroscopy.  A design of the experimental setup is described.  In Section~\ref{sec:LineShape}, the linear susceptibility and the Doppler-broadened line shapes are calculated.  

\section{Description of the Experimental Schemes}\label{sec:Experiment}
\begin{figure}[htbp]
\begin{center}
\includegraphics[scale = 0.26]{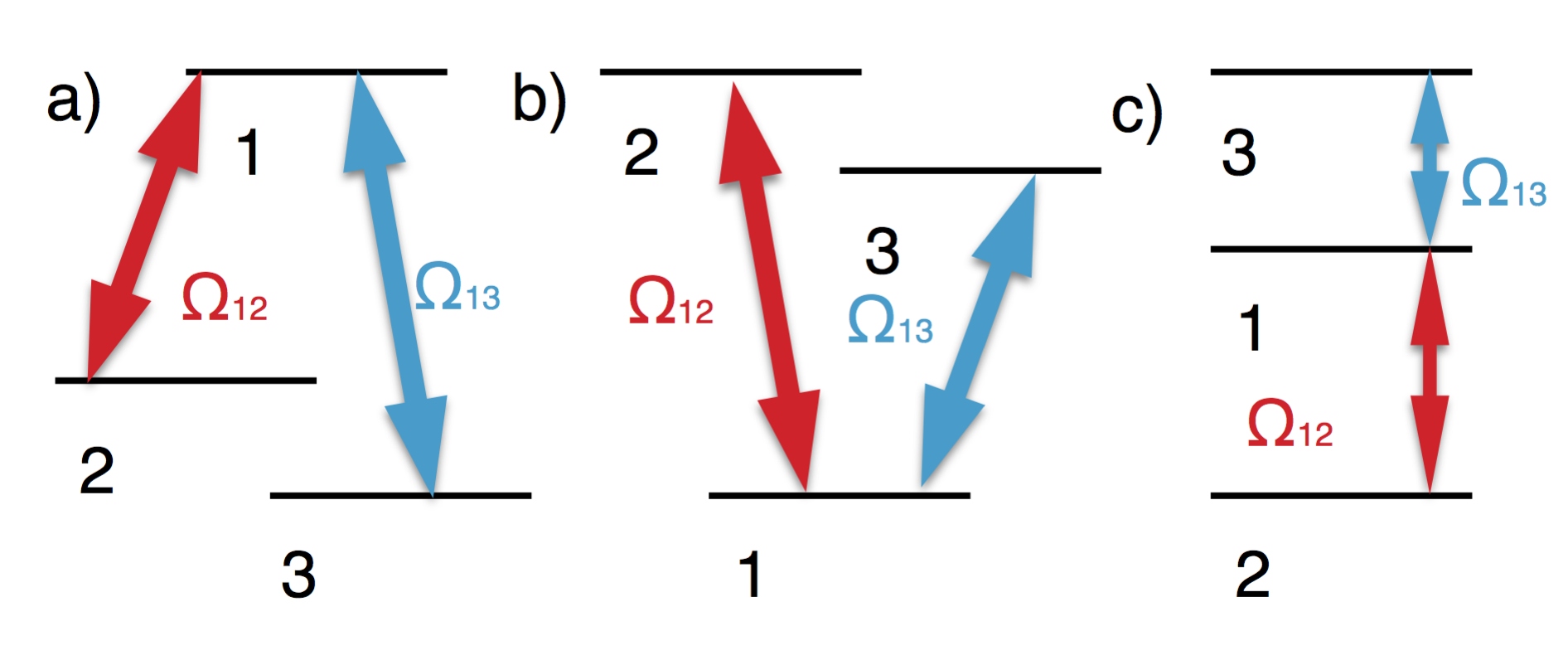}
\caption{\label{fig:LevelScheme} The level schemes of three-level systems. The red arrows represent the transition $P$ driven by the pump and probe beams.  The blue arrows represent the transition $C$ driven by the coupling beam. (a) $\Lambda$-scheme. (b) $V$-scheme. (c) Ladder scheme (or cascade scheme).} 
\end{center}
\end{figure}

The quantum coherence plays an important role of enhancing the coupled-resonance spectroscopy signal.  So one need to directly couple two transitions.  Fig.~\ref{fig:LevelScheme} shows three most common schemes of the three-level systems (ThLS).  They are usually referred to as (a) $\Lambda$-scheme, (b) $V$-scheme and (c) ladder scheme or cascade scheme. The EIT in these level schemes in various atoms has been reported~\cite{Fulton1995,Fleischhauer2005,Yang2011}.
An EIT experiment for ThLS usually involves at least two beams, the probe beam ($S$) and the coupling beam ($C$).   Conventional EIT spectra usually have Doppler-background that reduces the spectroscopic resolution.  Therefore EIT is not suitable for precision measurements or laser stabilization directly.

On the other hand, the Doppler-free saturation spectroscopy usually involves a pump beam and a probe beam driving a transition $P$ of two-level system (TLS), where the absorption of the probe beam on resonance is suppressed as the same group of the atoms has been pumped to saturate the transition, which leads to a dip in the Doppler broadening background.
The coupled-resonance spectroscopy~\cite{Burd2014} can be used to indirectly measure the coupling transition $C$ by probing transition $P$.

In light of the EIT and the coupled-resonance spectroscopy, one may enhance the coupled-resonance spectroscopy signal by coherently coupling the transition $C$ to the Doppler-free spectroscopic system so as to get rid of the Doppler background in the EIT and conduct the spectroscopic measurements with the coupled-resonance spectroscopy signal enhanced by EIT.

Ref.\citenum{Barker2015} and \citenum{Norcia2016} reported observations of $\ ^3P_{2} \rightarrow \ ^3S_{1}$ ($707 \text{nm}$) transition of $\text{Sr}$ in a hollow cathode lamp and stabilized the $707\text{nm}$ laser to the transition.    Since both the states are excited states, the thermal populations of the states are small. The other transitions from the same upper state to the the states in the same lower fine structure manifold are even harder to measure as the spontaneous decay rates are smaller.
Ref.\citenum{Barker2015} and \citenum{Zhang2016} reported observation and measurement of the transition of $\ ^3P_{0} \rightarrow \ ^3S_{1}$ ($679 \text{nm}$) by coupling it to the $\ ^3P_{2} \rightarrow \ ^3S_{1}$ ($707 \text{nm}$), nevertheless the signal is very weak.

\begin{figure}[htbp]
\begin{center}
\includegraphics[scale = 0.3]{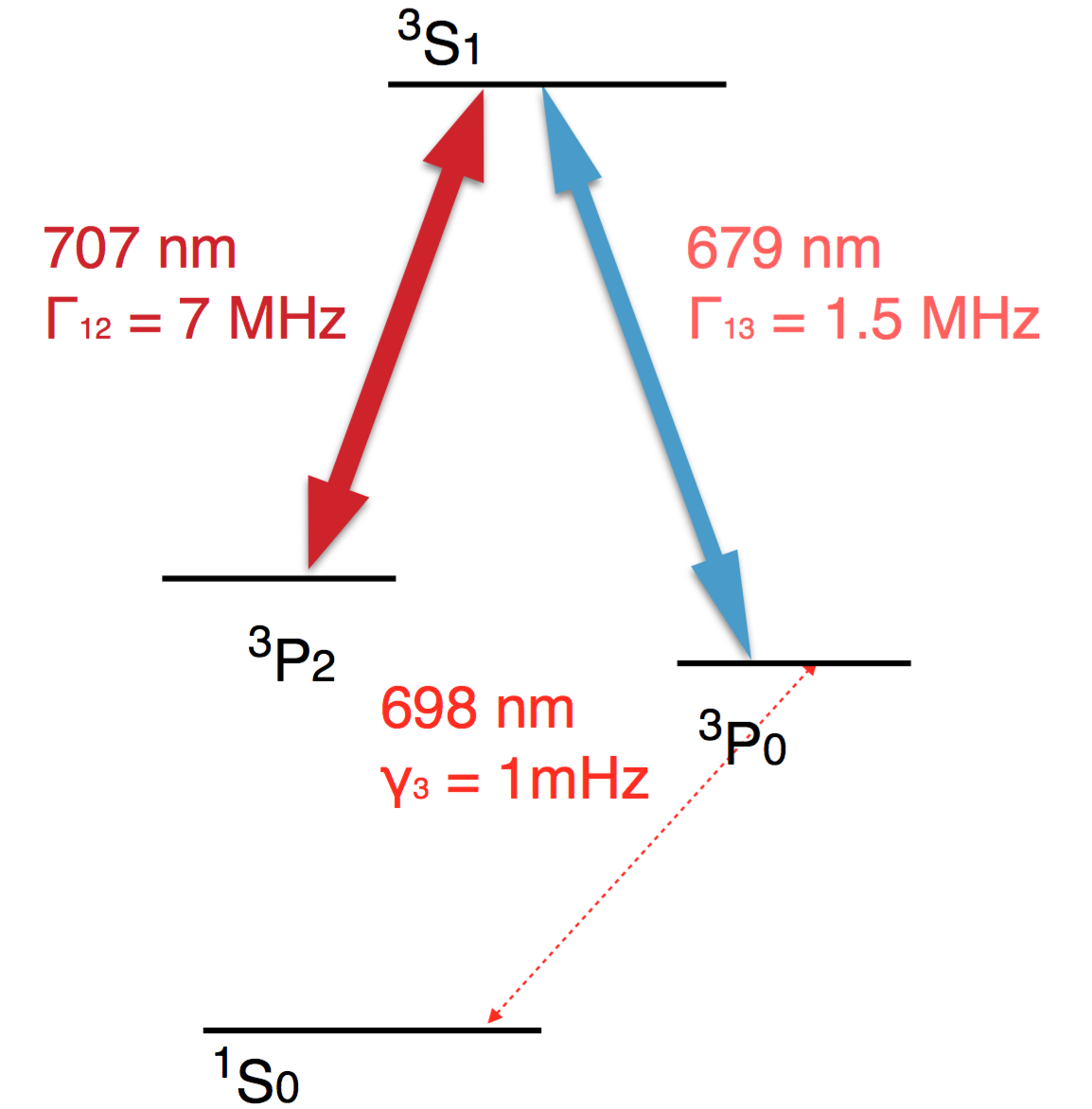}
\caption{\label{fig:EStructureSr} Relevant energy levels of $\text{Sr}$.  $\ ^3P_2$, $\ ^3P_0$ and $\ ^3S_1$ forms a ThLS of $\Lambda$-scheme.  The natural transition rate from $\ ^3P_0$ to  the ground state $\ ^1S_0$ is extremely small since this transition is electronic dipole forbidden.} 
\end{center}
\end{figure}

In this article we will mainly discuss about the ThLS of $\Lambda$-scheme. Particularly we will use the transitions between $\ ^3P_2$, $\ ^3P_0$ and $\ ^3S_1$ of the $\text{Sr}$ as an example.  The $V$- and ladder schemes are similar  while subjected to more restrictions in real applications.

The ThLS of $\Lambda$-scheme of $\text{Sr}$ may be ideal for EIT-enhanced coupled-resonance spectroscopy. Fig.~\ref{fig:LevelScheme} shows the relevant energy levels of the $\text{Sr}$. Comparing it to Fig.~\ref{fig:LevelScheme} a), we call the upper state $\ ^3S_1$ state $|1\rangle$,  the lower state $\ ^3P_2$ state $|2\rangle$,   and the lower state $\ ^3P_0$ state $|3\rangle$.  The transition between the states $|1\rangle$ and $|2\rangle$  is coupled by a pump beam and a probe beam as in the standard Doppler-free saturation spectroscopy.  We denote the frequency of the pump beam $\omega_p$, and its intensity $I_p$.  The detuning $\delta_p$ is defined as the difference of the beam frequency $\omega_p$ and the atomic transition frequency $\omega_{12}$, i.e. $\delta_p = \omega_p - \omega_{12}$. The Rabi frequency of the transition driven by the pump beam is denoted as $\Omega_p$.  All the physical quantities are defined for the probe beam similarly and are denoted by the same symbols with the suffix $p$ substituted by $s$.  The natural transition rate from the state $|1\rangle $ to the state $|2\rangle$ is $\Gamma_{12}$ due to the spontaneous emission.

A coupling beam of frequency $\omega_c$ and intensity $I_c$ is used to drive the transition between the state $|1\rangle$ and $|3 \rangle$, so the detuning is defined as $\delta_c = \omega_c - \omega_{13}$ and its Rabi frequency is denoted by $\Omega_c$.
It is noteworthy that since the transition from $\ ^3P_0$ to $\ ^1S_0$ is electronic dipole forbidden, the decoherence of the EIT coupling~\cite{Fleischhauer2005} is dominated only by collisions.

For the $V$- and ladder schemes, it is also possible to implement the EIT-enhanced coupled-resonance spectroscopy.
For example, $6S_{1/2} - 6P_{3/2}-8S_{1/2}$ transitions of cesium atom~\cite{Yang2011} and $5S_{1/2} - 5P_{3/2}-4D_{5/2}$ transitions of rubidium atom~\cite{Moon2007}.  Nevertheless, these schemes have some  disadvantages compared to the $\Lambda$-scheme.  
First, in the $V$- and ladder schemes, there is no EIT in the strict sense because of the absence of the (meta)stable dark state~\cite{Fleischhauer2005}.  
Second, in  the $V$- and ladder schemes, the state $|3\rangle$ is an excited state relative to the state $|1\rangle$, so there might be more decoherence processes involved.  This disadvantage can be saved by applying strong enough coupling beam.  
For the $V$-scheme, the two transitions share the same lower state, so either both of them can be measured easily or can hardly be measured due to the lack of thermal population if the  lower state is also an excited state.
So in the rest part, we will focus on the $\Lambda$-scheme.  For the $V$- and ladder scheme, the method can be generalized immediately as long as one follows the definitions and conventions above. 

The experimental setup is similar to the design reported in Ref.~\citenum{Zhang2016}.
The pump and probe beams are configured in the same way as the standard Doppler-free saturation spectroscopy.  The probe beam is modulated so that the absorption spectroscopy signal can be extracted by demodulation and so that  an error signal can be generated for locking the laser.
To implement the EIT-enhanced coupled-resonance spectroscopy, one need to tune the frequency and the intensity of the lasers properly so as to induce the coherent coupling.
It is also important that the coupling beam should be set to co-propagate with the probe beam, which eliminates the Doppler-shift of the two-photon detuning in EIT.  
The coupling beam is modulated independently so that the spectroscopy signal of the coupling transition $C$ can be separated from that of the transition $P$ and can be feedback to stabilize the coupling laser. 
The EIT will significantly enhance the weak coupled-resonance spectroscopy signal. 
With this experimental design we are able to lock two lasers simultaneously to one atomic source with the EIT-enhanced coupled-resonance spectroscopy.

\section{Theory of the Line Shape}\label{sec:LineShape}
The quantum interference and coherence play important roles in EIT-enhanced coupled-resonance spectroscopy in contrast to the conventional coupled-resonance spectroscopy.  To fully take the quantum coherence into account, the optical Bloch equations (OBEs) are solved instead of the rate equations.  The decoherence of the coupling transition is taken into account by adding the effective dephasing terms to the OBEs. The spontaneous emissions and dephasings cause the homogeneous broadening and lead to Lorentzian-like line shape functions. 

The imaginary part of the linear susceptibility $\chi^{(1)}(\omega; \{\delta_\alpha\}), \alpha =p,s,c$ at frequency $\omega$ gives the absorption of the beam.    
The line shape profile with  inhomogeneous Doppler broadening can be obtained by averaging the imaginary part of the linear susceptibility $\chi^{(1)}(\omega)$ over the thermal distribution of the atoms
\begin{equation}\label{eq:LineShape_LinearSuceptibilityDopplerBroadening}
\begin{split}
 \overline {\text{Im} \chi_s^{(1)} }&(\omega_s)=  \sqrt{\frac{m}{2\pi k_B T}} \int  \mathrm{d}v  \mathrm{e}^{-\frac{m v^2}{2k_BT}}  \text{Im} \chi_s^{(1)}(\omega_s; \delta'_p)  ,
\end{split}
\end{equation}
where the Gaussian factor $G(v) = \sqrt{\frac{m}{2\pi k_B T}} \mathrm{e}^{-\frac{m v^2}{2k_BT}}$ is the Gibbs distribution of the atoms and $\delta'_\alpha = \delta_\alpha \pm k_{\alpha} v$ is the effective detuning including the Doppler-shift seen by the atoms.
The integral in Eq.~\ref{eq:LineShape_LinearSuceptibilityDopplerBroadening} is calculated numerically.

Without loss of generality, let us consider the $\Lambda$-scheme defined above. The suffices $p$, $s$ and $c$ represent the pump, probe and coupling respectively.   The symbols are the same as they are defined in the Section~\ref{sec:Experiment}.  
The Hamiltonian of the ThLS under rotating wave approximation is 
\begin{equation} \label{eq:LineShape_ThLS_Hamiltonian}
\begin{split}
\mathcal{H}_{I}(t) = -\frac{\hbar}{2}
\begin{pmatrix}
0  &   \Omega_{12}(t)  &   \Omega_{13}(t)\\
\Omega_{12}^*(t)  & 0  & 0\\
\Omega_{13}^*(t) &0&0
\end{pmatrix},
\end{split}
\end{equation}
where $\Omega_{12}(t) = \Omega_p \mathrm{e}^{-\mathrm{i} \delta_p t } + \Omega_s \mathrm{e}^{-\mathrm{i} \delta_s t } $ and $\Omega_{13} = \Omega_c \mathrm{e}^{-\mathrm{i} \delta_c t }$.  This time  dependent Hamiltonian cannot be solved easily.

There are two regimes of EIT that concerns us. 
In the case of $|\Omega_p| \gg |\Omega_c|, |\Omega_s|$, the dependence on $\Omega_p$ can be adiabatically eliminated~\cite{Moler1992}.  The time dependence of the reduced Hamiltonian can be eliminated by going to the proper co-rotating frame, then we have the reduced effective
\begin{equation} \label{eq:LineShape_ThLS_Hamiltonian_predA}
\begin{split}
\mathcal{H}_{A}^{\text{red}} = -\frac{\hbar}{2}
\begin{pmatrix}
0  &   \Omega_{s}  &   \Omega_c \\
\Omega_{s}^* & -2\delta_s & 0\\
\Omega_c^*  &0& -2\delta_c
\end{pmatrix}.
\end{split}
\end{equation} 
As the price, we need to add the coherent pump terms to the equations of motion for the density matrix.

In the case of $|\Omega_c| \gg |\Omega_p|\gg |\Omega_s|$, the probe beam can be considered as a perturbation.
So we may first neglect the probe beam and solve the equations of motion and then do the perturbation.
By going to the proper co-rotating frame, then we have the time independent Hamiltonian
\begin{equation} \label{eq:LineShape_ThLS_Hamiltonian_predB}
\begin{split}
\mathcal{H}_{B}^{\text{red}} = -\frac{\hbar}{2}
\begin{pmatrix}
0  &   \Omega_{p}  &   \Omega_c \\
\Omega_{p}^* & -2\delta_p & 0\\
\Omega_c^*  &0& -2\delta_c
\end{pmatrix}.
\end{split}
\end{equation} 

The OBEs are given by $\dot{\rho} = \frac{1}{\mathrm{i} \hbar } [\mathcal{H} ,\rho ] + \mathcal{L}[\rho] + \mathcal{O}[\rho]$. The first term is the unitary evolution under the hermitian Hamiltonian $\mathcal{H}$.   The second term is the Lindblad superoperator, which includes the spontaneous emission and the dephasing terms.  The third operator consists of the other effects, like the collision redistribution, reduced coherent pumping, etc.

In the regime $|\Omega_p| \gg |\Omega_c| \gg |\Omega_s|$,  the $\Omega_p$ can be adiabatically eliminated, so we have the reduced OBEs
\begin{align} % \label{eq:LineShape_ThLS_OBEs}
\begin{split}\label{eq:LineShape_ThLS_OBEs_a}
\dot{\rho}_{11} =& \frac{\mathrm{i}}{2} \Big( \Omega_s \rho_{21}- \rho_{12} \Omega_s^*  +
\Omega_c \rho_{31}- \rho_{13} \Omega_c^*   \Big)\\
 &-  \Gamma_{12}   \rho_{11} - \Gamma_{13}   \rho_{11} 
 + \Gamma_p (\rho_{22} - \rho_{11}),
\end{split}\\
\begin{split}\label{eq:LineShape_ThLS_OBEs_b}
\dot{\rho}_{22} =& \frac{\mathrm{i}}{2} \Big( \Omega_s^* \rho_{12}- \rho_{21} \Omega_s   \Big) +\Gamma_{12} \rho_{11} \\
&+ R(\rho_{33} -\rho_{22}) 
+ \Gamma_p (\rho_{11} - \rho_{22}) ,
\end{split}\\
\begin{split}\label{eq:LineShape_ThLS_OBEs_c}
\dot{\rho}_{33} =& \frac{\mathrm{i}}{2} \Big( \Omega_c^* \rho_{13}- \rho_{31} \Omega_c   \Big)
+ \Gamma_{13} \rho_{11}  + R(\rho_{22} -\rho_{33})  ,
\end{split}\\
\begin{split}\label{eq:LineShape_ThLS_OBEs_d}
\dot{\rho}_{12} =&\frac{\mathrm{i}}{2} \Big( \Omega_s \rho_{22} +  \Omega_c \rho_{32}  - \rho_{11} \Omega_s + \rho_{12} 2\delta'_s \Big) \\
& - \frac{1}{2} \Gamma_{12}   \rho_{12} - \frac{1}{2} \Gamma_{13}   \rho_{12}
 -  \frac{1}{2} \gamma_{1}   \rho_{12}-  \frac{1}{2} \gamma_{2}   \rho_{12},
\end{split}\\
\begin{split}\label{eq:LineShape_ThLS_OBEs_e}
\dot{\rho}_{13} =&\frac{\mathrm{i}}{2} \Big( \Omega_s \rho_{23} +  \Omega_c \rho_{33}  - \rho_{11} \Omega_c + \rho_{13} 2\delta'_c \Big)\\
& - \frac{1}{2} \Gamma_{12}   \rho_{13}  - \frac{1}{2}  \Gamma_{13}   \rho_{13} 
 -  \frac{1}{2} \gamma_{3}   \rho_{13}  -  \frac{1}{2}\gamma_{1}   \rho_{13} ,
\end{split}\\
\begin{split}\label{eq:LineShape_ThLS_OBEs_f}
\dot{\rho}_{23} =&\frac{\mathrm{i}}{2} \Big( \Omega_s^* \rho_{13} -  2\delta'_s \rho_{23}  - \rho_{21} \Omega_c + \rho_{23} 2\delta'_c  \Big)\\
&-  \frac{1}{2}  \gamma_{3}   \rho_{23} -  \frac{1}{2}  \gamma_{2}   \rho_{23},
\end{split}
\end{align}
where $\Gamma_p = \frac{|\Omega_p|^2}{4(\delta'_p)^2 + \Gamma_{12}^2} \Gamma_{12} = \frac{I_p}{2I_p^{\text{sat}}(\omega_p)} \Gamma_{12} $ is the coherent pumping rate,
$\gamma_i, i=1,2,3$ are the dephasing rates from the term $\sum_i \frac{1}{2} \gamma_i (2 \hat{\sigma}_{ii}\rho \hat{\sigma}_{ii} - \hat{\sigma}_{ii} \rho-\rho \hat{\sigma}_{ii})$ describing the energy-conserving dephasing processes~\cite{Fleischhauer2005},
and $R$ is (non-radiative decay) population transfer rate between the lower states~\cite{Burd2014}.
Since $|\Omega_c| \gg |\Omega_s|$, Eq.~\ref{eq:LineShape_ThLS_OBEs_f} gives
\begin{equation}\label{eq:LineShape_ThLS_OBEs_f_sol}
\begin{split}
\bar{\rho}_{23} =& \frac{-\Omega_c}{ 2\delta'_s   -2\delta'_c -\mathrm{i}  (\gamma_{3} + \gamma_{2})  }\rho_{21}  
= \frac{-\Omega_c}{ 2\Delta'  -\mathrm{i} \gamma_{23}  }\rho_{21}  ,
\end{split}
\end{equation}
where $\Delta' = \delta'_s   -\delta'_c$ is the effective two-photon detuning and $\gamma_{23} = \gamma_2 + \gamma_3$.  Substituing Eq.~\ref{eq:LineShape_ThLS_OBEs_f_sol} for the $\rho_{23}$ in Eq.~\ref{eq:LineShape_ThLS_OBEs_d}, one obtains 
\begin{equation}\label{eq:LineShape_ThLS_OBEs_d_sol}
\begin{split}
\bar{\rho}_{12,s} =& - \frac{\Omega_s (\rho_{22} - \rho_{11})}{ 2\delta'_s  - \frac{|\Omega_c|^2}{2\Delta' + \mathrm{i} \gamma_{23} }  + \mathrm{i}  \gamma_{12}}, 
\end{split}
\end{equation}
where $ \gamma_{12} = \Gamma_{12}  + \Gamma_{13}  +\gamma_1 + \gamma_2$.

\begin{figure}[htbp]
\begin{center}
\includegraphics[scale = 0.54]{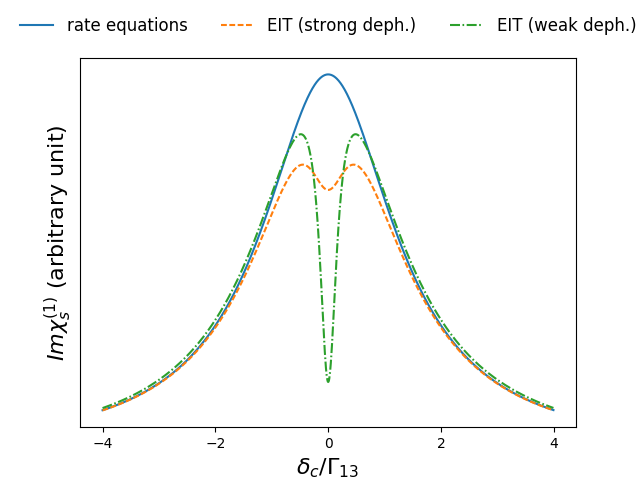}
\caption{\label{fig:ImgChi1_s} Imaginary part of the susceptibility $\text{Im} \chi_s^{(1)}$.  The solid blue line shows the susceptibility calculated by rate equations.  The green dash-dot line shows the susceptibility with weak dephasing $\gamma_2 = \gamma_3 =0.05$. The yellow dashed line shows the susceptibility with dephasing $\gamma_2 = \gamma_3 =0.5$. The parameters for this plot are $\Gamma_{12} = 7, \Gamma_{13} = 1.5, \gamma_1 = 0.1, \Omega_p =7, \Omega_c = 2, \Omega_s =0.1, \delta_p = \delta_s =0$.  } 
\end{center}
\end{figure}

So the linear susceptibility for the probe beam at $\omega_s$ is
\begin{equation}\label{eq:LineShape_ThLS_LinearSusceptibility}
\begin{split}
\chi^{(1)}_{s}(\omega_s) =& - \frac{4\pi |d_{12}|^2 n}{\hbar} \frac{(\rho_{22} - \rho_{11})(2\Delta' + \mathrm{i} \gamma_{23} )}{ -|\Omega_c|^2 +(2\Delta' + \mathrm{i} \gamma_{23} ) (2\delta'_s  +  \mathrm{i}  \gamma_{12})}\\
=& - \frac{4\pi |d_{12}|^2 n}{\hbar} \frac{(\rho_{22} - \rho_{11})}{ \left| |\Omega_c|^2 -(2\Delta' + \mathrm{i} \gamma_{23} ) (2\delta'_s  +  \mathrm{i}  \gamma_{12}) \right|^2}\\
&\times \Big[ ( -2\Delta' (|\Omega_c|^2 - 4\Delta' \delta'_c) +2\delta'_c  \gamma_{23}^2) \\
&- \mathrm{i}(4(\Delta')^2 \gamma_{12} + \gamma_{23}(|\Omega_c|^2 + \gamma_{23}\gamma_{12}) )\Big] , 
\end{split}
\end{equation}
where $n$ is the number density of the atoms.
The population distribution approximately takes the value predicted by the rate equations.
In the ideal case $\gamma_{23} = 0$,  in contrast to the result of the rate equations, $\chi^{(1)}_{s}(\omega_s)$ vanishes, which is the signature of EIT.   
In the practice, the dephasing terms may cause the decoherence.  Nevertheless, as long as  $|\Omega_c|^2 \gg \gamma_{23} \gamma_{12}$, the EIT features should still be  observable~\cite{Fleischhauer2005}.

Fig.~\ref{fig:ImgChi1_s} shows the dependence of  $\text{Im} \chi_s^{(1)}$ on $\delta_c$, when pump and probe beams are on resonance ($\delta_p =\delta_s = 0$).  The result of rate equations differs significantly from EIT results.  When the dephasing is weak $\gamma_2 = \gamma_3 =0.05$, the central dip indicates a significant suppression of absorption.  When the dephasing is an order of magnitude stronger $\gamma_2 = \gamma_3 =0.5$, although the dip is much shallower,  the suppression of absorption is not negligible.  In $\text{Sr}$, the two lower levels have very narrow natural line with, the dephasing is dominated by the collisions.

\begin{figure}[htbp]
\begin{center}
\includegraphics[scale = 0.62]{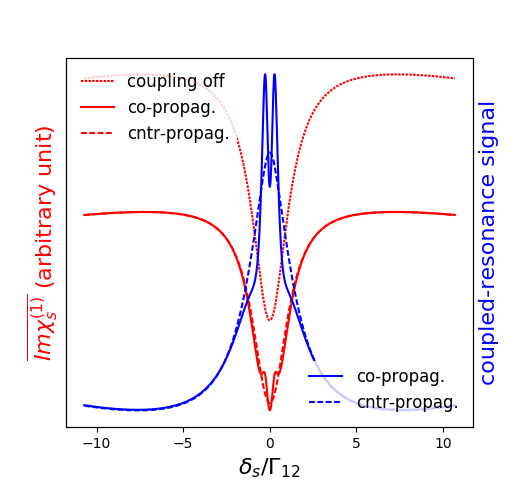}
\caption{\label{fig:Doppler_ImgChi1_s} Doppler-broadened  absorption and coupled-resonance signal in the regime $|\Omega_p| \gg |\Omega_c| \gg |\Omega_s|$. 
(a) The red lines are Doppler-broadened absorption curves.  The red dotted line is the absorption curve when coupling beam is turned off.  The Lamb dip is power-broadened by the pump beam.  The red solid and dashed line are the Doppler-broadened absorption curves  when the coupling beam is co-propagating and counter-propagating with the probe beam respectively.   (b)  The blue solid and dashed line are the EIT-enhanced coupled-resonance signals for the co-propagating and counter-propagating configurations respectively.   The parameters for the plot are $\Gamma_{12} = 7, \Gamma_{13} = 1.5, \gamma_1 = 0.1, \Omega_p =16, \Omega_c = 8, \Omega_s =0.1, \delta_c =0$. The collision-dominated dephasing terms are $\gamma_2 = \gamma_3 =1.5$ and  non-radiative decay transition rate $R = 3.0$ is estimated from Ref.~\citenum{Zhang2016}.  }
\end{center}
\end{figure}

The Doppler-broadened absorption is calculated according to  Eq.~\ref{eq:LineShape_LinearSuceptibilityDopplerBroadening}.  Fig,~\ref{fig:Doppler_ImgChi1_s} shows the Doppler-broadened  absorption and coupled-resonance signal in the regime $|\Omega_p| \gg |\Omega_c| \gg |\Omega_s|$.
The red solid and dashed line are the Doppler-broadened absorption curves  when the coupling beam is co-propagating and counter-propagating with the probe beam respectively. 
 Both of them has a significant suppression of the absorption due to the EIT (compared to red dotted line for the coupling beam turned off). The sharper dip of the red solid line on top of the Lamb dip is particularly due to the EIT-enhancement.  
For the coupling beam co-propagates with the probe beam, the Doppler shift for the two-photon detuning is partially cancelled. (For $\text{Sr}$, the two-photon detuning has a Doppler shift $v_T(k_s-k_c) \approx 0.04 k_s v_T$.)  For the coupling beam counter-propagates with the probe beam, the Doppler shift doubles the two-photon detuning. 
The difference of the absorption when the coupling beam is on and off give the EIT-enhanced coupled-resonance signal.  In the co-propagating configuration, the coupled-resonance signal has a sharp EIT-enhanced dip, while the counter-propagating configuration the EIT-enhanced dip is rounded-off by the two-photon Doppler-shift.

In the regime $|\Omega_c| \gg |\Omega_p| \gg |\Omega_s|$, the $\omega_s$ can be considered as a perturbation, 
so one obtains OBEs very similar to the Eq.~\ref{eq:LineShape_ThLS_OBEs_a} to Eq.~\ref{eq:LineShape_ThLS_OBEs_f}, but without coherent pumping term $\Gamma_p(\rho_{11}-\rho_{22})$ and with all the suffix $s$ substituted by $p$.
One obtains the steady state solutions
\begin{equation}\label{eq:eq:LineShape_ThLS_OBEs_sol1_cp}
\begin{split}
\bar{\rho}_{23}  =& \frac{-\Omega_c}{ 2\tilde{\Delta}'  -\mathrm{i} \gamma_{23}  }\bar{\rho}_{21}  ,\\
\bar{\rho}_{12} =&  -\frac{\Omega_p}{2\delta'_p + \mathrm{i} \gamma_{12}}  -  \frac{\Omega_c}{2\delta'_p + \mathrm{i} \gamma_{12}}  \bar{\rho}_{32}  \\
=& -\frac{\Omega_p}{2\delta'_p + \mathrm{i} \gamma_{12}} \frac{1}{\left(1-  \frac{|\Omega_c|^2}{(2\delta'_p + \mathrm{i} \gamma_{12})(2\tilde{\Delta}'  + \mathrm{i} \gamma_{23} )} \right)}\\
\bar{\rho}_{13} =&- \frac{\Omega_p}{2\delta'_c+\mathrm{i}  \gamma_{12}   }\bar{\rho}_{23} -  \frac{\Omega_c( \rho_{33}  - \rho_{11} )}{2\delta'_c+\mathrm{i}  \gamma_{12}   }  \\
=& -\frac{\Omega_c}{2\delta'_c+\mathrm{i}  \gamma_{12}} \Bigg(( \rho_{33}  - \rho_{11} )\\
&+ \frac{|\Omega_p|^2}{(2\delta'_p - \mathrm{i} \gamma_{12})  ( 2\tilde{\Delta}'  -\mathrm{i} \gamma_{23})   - |\Omega_c|^2 } \Bigg)
\end{split}
\end{equation}
where  $\tilde{\Delta}'  = \delta'_p - \delta'_c$ is the effective two-photon detuning.
The solution for $\rho_{23}$ is formally similar to Eq.~\ref{eq:LineShape_ThLS_OBEs_f_sol}.
The factor $\frac{1}{1-  \frac{|\Omega_c|^2}{(2\delta'_p + \mathrm{i} \gamma_{12})(2\Delta'  + \mathrm{i} \gamma_{23} )} }$ in the solution for $\rho_{12}$ containing essential features of EIT modifies the solution of $\rho_{12}$ to the TLS, which suppress the absorption of pump beam on resonance.

The perturbation in terms of $\Omega_s$ leads the component of $\rho_{12}$ at $\omega_{s}$
\begin{equation}\label{eq:eq:LineShape_ThLS_OBEs_sol2_cps}
\begin{split}
\bar{\rho}_{12,s} =& -\frac{\Omega_s}{2\delta'_s + \mathrm{i} \gamma_{12}} \frac{1}{\left(1-  \frac{|\Omega_c|^2}{(2\delta'_s + \mathrm{i} \gamma_{12})(2\Delta'  + \mathrm{i} \gamma_{23} )} \right)}
\end{split}
\end{equation}
where the two-photon detuning $\Delta' = \delta'_s - \delta'_c$.
This result agrees with Eq.~\ref{eq:LineShape_ThLS_OBEs_d_sol} in the limit of $\rho_{22}\approx 1$.
Another necessary condition for this approximation to be valid is the transition rate $R$ to be much smaller than $\Gamma_p$ and $\Gamma_c$, which is implied by the EIT condition $|\Omega_c|^2 \gg \gamma_{23} \gamma_{12}$.

\begin{figure}[htbp]
\begin{center}
\includegraphics[scale = 0.62]{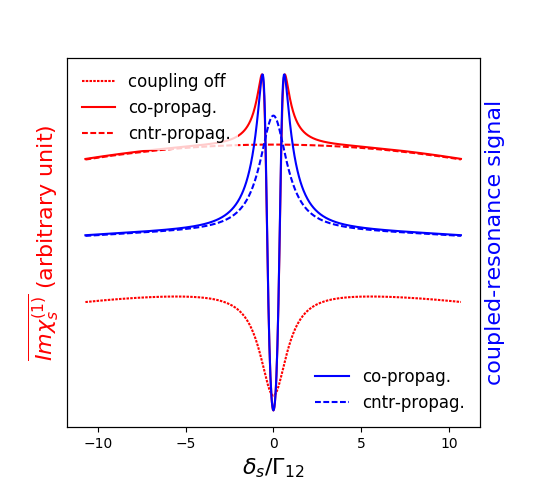}
\caption{\label{fig:Doppler_ImgChi2_s} Doppler-broadened  absorption and coupled-resonance signal in the regime $|\Omega_c| \gg |\Omega_p| \gg |\Omega_s|$. 
(a) The red lines are Doppler-broadened absorption curves.  The red dotted line is the absorption curve when coupling beam is turned off.  The red solid and dashed line are the Doppler-broadened absorption curves  when the coupling beam is co-propagating and counter-propagating with the probe beam respectively.   (b)  The blue solid and dashed line are the EIT-enhanced coupled-resonance signal for the co-propagating and counter-propagating configurations respectively.   The parameters for the plot are the same as in the Fig.~\ref{fig:Doppler_ImgChi1_s} but $\Omega_p =8, \Omega_c = 16$.  }
\end{center}
\end{figure}

Fig.~\ref{fig:Doppler_ImgChi2_s} shows the Doppler-broadened  absorption and coupled-resonance signal in the regime $|\Omega_c| \gg |\Omega_p| \gg |\Omega_s|$.  The overall absorption is larger when coupling beam is on as as the atoms are pumped to the state $|2\rangle$, so the Doppler-free saturation dip vanishes.  When the coupling beam counter-propagates with the probe beam, the EIT dip is smeared  out by the two-photon Doppler shift.  When the coupling beam is co-propagates with the probe beam, it shows an Dopper-free EIT dip.

\section{Conclusion}\label{sec:conclusion}
We have presented proposal of the EIT-enhanced coupled-resonance spectroscopy.   The line shape has been calculated by solving the OBEs.  The results differ from the prediction of rate equations in a regime where coherence and quantum interference are not negligible.  In both regime, the EIT feature significantly enhances coupled-resonance spectroscopy signal and leads to a sharp resonance peak.  Furthermore, with the EIT-enhanced coupled-resonance spectroscopy, we may build a compact setup for locking two lasers simultaneously to one atomic source.
   
\section{Acknowledgement}
The author would acknowledge G.~K.~Campbell, D.~Barker, N.~Pisenti and B.~Reschovsky for helpful discussions.   
Part of this work is done in Joint Quantum Institute, National Institute of Standards and Technology, and University of Maryland.

\bibliography{CRSref}
\end{document}